\documentclass[12pt,preprint]{aastex}

\newcommand\gele{G11.2--0.3}%
\newcommand\kms{$\rm km\;s^{\rm -1}$}%
\newcommand\cms{$\rm cm^{\rm -3}$}%
\newcommand\um{$\mu$m}%
\newcommand{\feii}{[\ion{Fe}{2}]}%
\newcommand{\arich}{$\alpha$-rich freezeout}%
\newcommand{\sii}{[\ion{S}{2}]}%
\newcommand{\si}{[\ion{S}{1}]}%
\newcommand{\ab}{A,~B}%
\newcommand{\cde}{C,~D,~E}%
\def\simgt{\lower.5ex\hbox{$\; \buildrel > \over \sim \;$}}
\def\simlt{\lower.5ex\hbox{$\; \buildrel < \over \sim \;$}}

\slugcomment{\today}

\shorttitle{Iron Ejecta and Supernova Explosion in G11.2--0.3}
\shortauthors{Moon et al.}

\begin{document}

\title{Dense Iron Ejecta and Core-collapse Supernova Explosion in the Young Supernova Remnant G11.2--0.3\footnote{Based in part on data collected at Subaru Telescope, which is operated by the National Astronomical Observatory of Japan.}}

\author{Dae-Sik~Moon\altaffilmark{1},
Bon-Chul~Koo\altaffilmark{2},
Ho-Gyu~Lee\altaffilmark{3},
Keith~Matthews\altaffilmark{4},
Jae-Joon~Lee\altaffilmark{5},
Tae-Soo~Pyo\altaffilmark{6} 
Ji~Yeon~Seok\altaffilmark{2} 
Masahiko~Hayashi\altaffilmark{6} }
\altaffiltext{1}{Department of Astronomy and Astrophysics, University of Toronto, Toronto, ON M5S 3H4, Canada; moon@astro.utoronto.ca}
\altaffiltext{2}{School of Physics and Astronomy, Seoul National University, Seoul, 151-742, Korea}
\altaffiltext{3}{Department of Astronomy, Graduate School of Science, The University of Tokyo, Bunkyo-ku, Tokyo, 113-0033, Japan}
\altaffiltext{4}{Division of Physics, Mathematics and Astronomy, California Institute of Technology, MS 320-47, Pasadena, CA 91125}
\altaffiltext{5}{Department of Astronomy and Astrophysics, Pennsylvania State University, University Park, PA 16802}
\altaffiltext{6}{Subaru Telescope, National Astronomical Observatory of Japan, 650 North Aohoku Place, Hilo, HI 96720, USA}

\begin{abstract}
We present the results of near-infrared spectroscopic 
observations of dense ($\simgt$ 10$^3$~cm$^{-3}$) 
iron ejecta in the young core-collapse supernova remnant \gele. 
Five ejecta knots projected to be close to its center
show a large dispersion in their Doppler shifts:
two knots in the east are blueshifted by more than 1,000 \kms, 
while three western knots have relatively small blueshifts of 20--60 \kms.
This velocity discrepancy may indicate that the western knots
have been significantly decelerated
or that there exists a systematic velocity difference among the knots.
One ejecta filament in the northwestern boundary, on the other hand, 
is redshifted by $\simgt$~200~\kms, while opposite filament in 
the southeastern boundary shows a negligible radial motion.
Some of the knots and filaments have secondary velocity components,
and one knot shows a bow shock-like feature in the velocicty structure.
The iron ejecta appear to be devoid of strong emission from other heavy elements, 
such as S,
which may attest to the \arich\ process in the explosive nucleosynthesis 
of the core-collapse supernova explosion close to its center.
The prominent bipolar distribution of the Fe ejecta in the 
northwestern and southeastern direction, along with the elongation of
the central pulsar wind nebula in the perpendicular direction,
is consistent with the interpretation that the supernova exploded 
primarily along the northwestern and southeastern direction.
\end{abstract}


\section{Introduction}

The final fatal scene of a massive ($\simgt$~8~$M_{\rm \odot}$) star
is a sudden collapse of its Fe core and a subsequent appearance of a supernova (SN).
The study of the core-collapse SNe has been one of the centerpieces 
of the modern astrophysics, accompanying various fundamental physics 
aided by the state-of-art numerical simulations and dedicated observations.
Although how the core-collapse SNe explode still needs to be understood 
despite extensive studies of the past and present,
one unequivocal prediction is the formation of Fe ejecta as the final product of 
the nucleosynthesis deep inside a progenitor.
Since the formation and expulsion of the Fe is very sensitive to the SN explosion details,
the dynamics and chemistry of the Fe ejecta in young core-collapse SN remnants (SNRs)
can in principle provide critical information to understanding the core-collapse SN explosion.
However, Fe ejecta in SNRs are elusive, 
with very limited detections to date concentrated on hot diffuse X-ray ejecta.
In the optical waveband, which is sensitive to dense, cold ejecta, 
all the observed ejecta are dominated by the O, N, Si, or S with little contribution from the Fe.
So far dense Fe-rich ejecta have never been detected as far as we are aware,
and we report here the first such case in \gele.

\gele\ is a young SNR thought to be the remains of the historical SN of AD~386 \citep{cs77}. 
It has a shell of 2$'$ radius at 5~kpc distance \citep{get88}, 
bright both in the radio and X-rays,
along with a central pulsar and its elongated wind nebula \citep{get88,ket01}.
\gele\ was proposed to be an evolved version of the Cassiopeia~A (Cas~A) SNR 
because of their similar radio shells \citep{get88},
and both may be a SN~IIL/b with a red supergiant progenitor \citep{che05}.
Recently we discovered bright near-infrared (IR) \feii\ emission 
from the Fe ejecta in \gele\ \citep[][hereafter Paper I]{ket07}.
The \feii\ emission concentrates on two opposite filaments at its 
northwestern (NW) and southeastern (SE) boundaries,
while numerous smaller clumpy knots scattered across the SNR.
Our high-resolution spectroscopic observations of the \feii\ emission
presented here reveal clear velocity differences among the Fe ejecta 
as well as its chemical composition, 
providing information for understanding the nature of the ejecta
in connection with the core-collapse SN explosion in \gele.

\section{Observations and Results}

We performed near-IR \feii\ line spectroscopic observations of the SNR \gele\ in 
2006 June 6--9 using the Long-slit Near-Infrared Spectrograph \citep{let96} of 
the Palomar 5-m Hale Telescope.
The long-slit spectrograph is equipped with a NICMOS 256 $\times$ 256 pixels HgCdTe detector,
and the slit length and width were 38\arcsec\ and 1\arcsec, respectively.
We first obtained high-resolution ($R$ $\simgt$ 4000) spectra of the \feii\ 1.644 $\mu$m line 
from five bright Fe knots (which we name knots A--E) around the central 
pulsar\footnote{The pulsar is located very close to the geometrical center of \gele\ \citep{ket01}, 
and we adopt the pulsar position as the central position of the SNR in this study.} and the NW and SE filaments,
measuring their radial velocities (Fig.~\ref{fig_overall}).
We then observed the knots and filaments with the low-resolution ($R$ $\simeq$ 700) 
mode to detect both \feii\ 1.644 and 1.599 $\mu$m lines simultaneously
and used their line intensities to calculate the electron number densities (see below).
We also observed He~I (1.083 $\mu$m) and \sii\ (1.029--1.037 $\mu$m) lines from the knots and filaments, 
but detected only the He~I line emission from the SE filament.
All the knots and filaments were observed at least twice dithered along the slit,
and the sky emission, including the OH sky lines, was subtracted using the dithered frame.
In addition, we observed the knot~A with the Infrared Camera and Spectrograph \citep[IRCS;][]{kobaet00} 
of the 8-m Subaru Telescope on 2006 July 25.
We obtained a spectrum covering almost the entire near-IR atmospheric windows of
$zJHK$ bands in the range of 1.03--2.49 \um\ using the echelle mode 
with two Aladdin III $1024\times 1024$ InSb arrays.
The slit width of IRCS was $0.''54$ which corresponds to the spectral resolution $R$ $\simeq$ 5,000.

Fig.~\ref{fig_overall} presents the Palomar near-IR spectra of the \feii\ 1.644~$\mu$m transition 
of five Fe knots projected to be around the center and two Fe filaments at the NW and SE boundaries of \gele\
overlaid on a \feii\ 1.644~$\mu$m line emission image (Paper~I).
The contours in the main left panel represent the VLA radio image \citep{get88}.
In the right panel, the thick solid contour delineates the boundary of the strong Chandra X-ray
emission of the pulsar wind nebula (PWN) elongated in the northeastern (NE) and southwestern (SW) direction,
while the red dashed contour represents
the boundary of the PWN from radio tomography \citep{ket01,trk02}.
The ejecta knots are projected to be within or close to the radio boundary of the PWN. 
The \feii\ spectra exhibit clear velocity differences among the knots and filaments: 
A, B in the east are blueshifted by $\simgt$ 1,000 \kms;
\cde\ in the west have relatively small blueshifts of $\simlt$~60~\kms. 
The NE filament, in contrast, shows redshifts of $\simgt$~200~\kms;
the SW filament has a small Doppler shift motion of --10 \kms.
The NW--SE direction of the two filaments is almost perpendicular to the
NE--SW elongation direction of the X-ray PWN.
We note that in the spectra of the knots C and D there appear to be weak emission components
around --1,000 \kms, which are mainly caused by the imperfect sky subtraction of the OH 
sky line at 1.6389 $\mu$m. The imperfect sky subtraction is also responsible for the 
negative dips adjacent to the emission components in the spectra. However, for the knot C,
in addition to the contribution from the imperfect sky subtraction, we find that there might be 
additional emission from the source itself around --1,000 \kms, although the intensity is too faint to be confirmed.
Similarly the weak emission component in the knot B around 0 \kms\ is also caused
by the imperfect sky subtraction of the OH sky line at 1.6442 $\mu$m.

\begin{figure}[htf]
\plotone{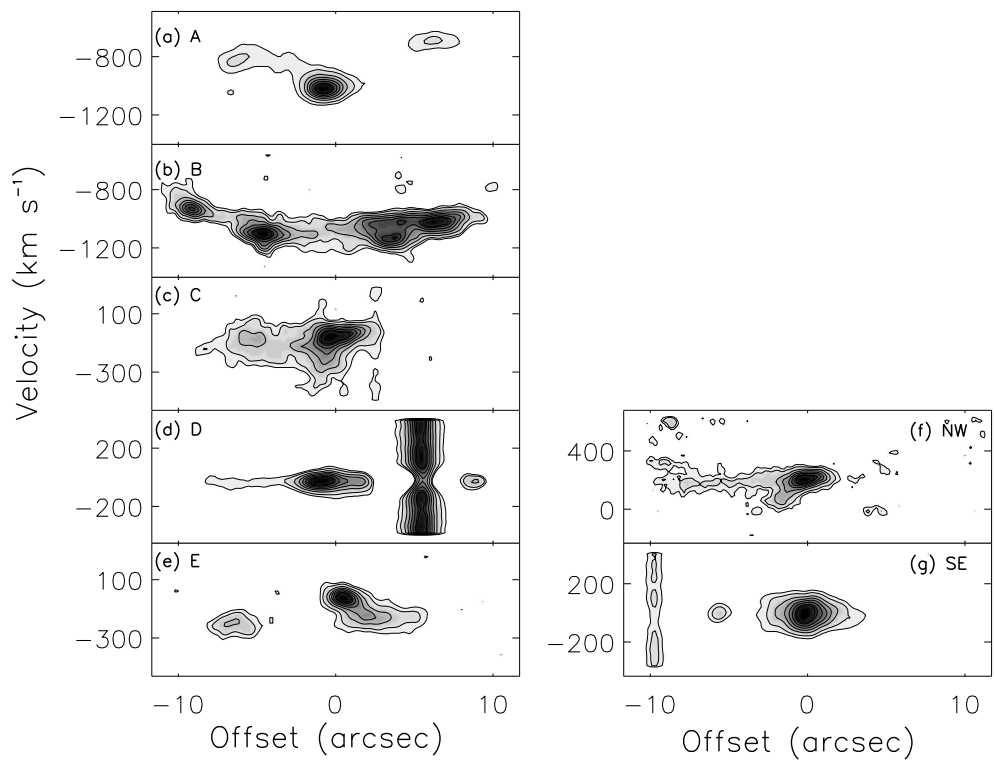}
\caption{
(a)--(g): P-V diagrams of the Fe emission of the ejecta knots and filaments.
The abscissa corresponds to the offset along the slit.
The continuum features in (d) and (g) are due to a nearby star.
For the knot C, there is no data coverage beyond the offset --10\arcsec.
Note that the eastern knot~A has weak secondary components redshifted with 
respect to the main component, while B may have multple components.
In contrast, the two western knots (C and E) and the NW filament
show weak secondary components blueshifted compared to the main component.
}
\label{fig_pv}
\end{figure}

The \feii\ 1.644~$\mu$m line spectra in Fig.~\ref{fig_overall} show deviations 
from a simple symmetric shape except for the SE filament.
It is apparent for B, C, and E with broad double-peak or flat-top profiles, 
while A, D, and the NW filament appear to have relatively weak secondary components
slightly deviating from a symmetric profile.
Fig.~\ref{fig_pv} compares their position-velocity (P-V) diagrams 
of the \feii\ 1.644~$\mu$m line emission,
revealing the velocity structures more clearly. 
A and B in the east have secondary velocity components
redshifted from the main component by $\simgt$~200~\kms;
to the contrary, C, E and the NW filament in the west have blueshifted secondary components.
For the knot A, its central part appears to have a bow shock-like feature in the P-V diagram (see below).
In Fig.~\ref{fig_pv2} we present the P-V diagrams of the entire velocity range of --1400 and 400 \kms\
covered by our observations,
highlighting negative pixels caused by the subtraction of a dithered frame by contours.
The aforementioned negative OH residuals are visible in some panels, especially in (d),
while there is no bright Fe emission outside the velocity ranges covered in Fig.~\ref{fig_pv}.

\begin{figure}[htf]
\plotone{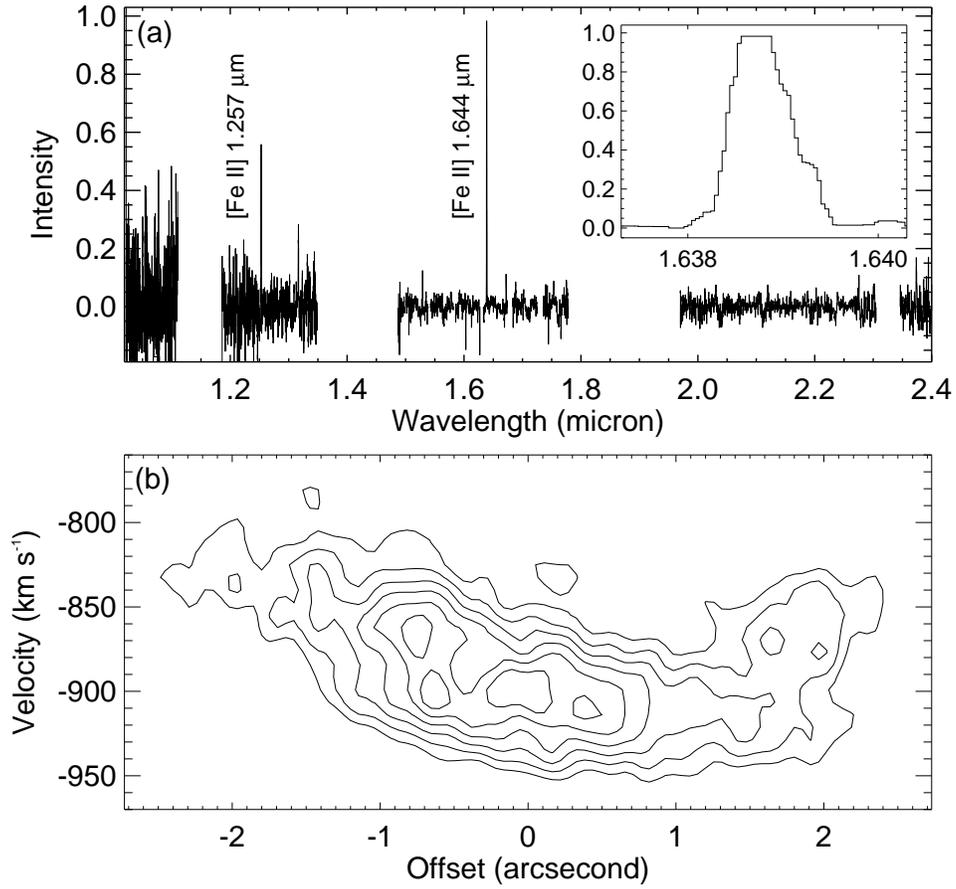}
\caption{
(a) Subaru near-IR spectrum of the ejecta knot~A in the 1.02--2.4~$\mu$m range. 
The intensity is normalized by the peak intensity of the \feii\ 1.644 $\mu$m line.
Only the lines from the iron ion (Fe$^+$) are detected.
The inset provides a magnified view of the 1.644~$\mu$m line.
(b) P-V diagram of the \feii\ 1.644 line which reveals a bow shock-like feature.
}
\label{fig_subaru}
\end{figure}

Fig.~\ref{fig_subaru} (top panel) presents our Subaru spectrum of the knot~A 
covering most of the near-IR atmospheric windows in the 1.02--2.4 $\mu$m range.
Only the lines from Fe ion (Fe$^+$) are detected, 
including the strong transitions at 1.257 and 1.644 $\mu$m 
and relatively weak ones at 1.321, 1.534, and 1.664 $\mu$m.
The spectrum is somewhat noisy at $\simlt$~1.1~$\mu$m, 
but appears to be devoid of the S (\sii\ $\simeq$ 1.290--1.037~$\mu$m and \si\ $\simeq$ 1.082~$\mu$m)
and O (O~I $\simeq$ 1.129~$\mu$m) lines which are much brighter than the Fe lines in 
the O-rich SN ejecta of Cas~A \citep{gf01}.
The upper limits, at the 90~\% confidence level, of the line intensities of these lines are
approximately 0.17 (\sii\ lines), 0.15 (\si\ line), and 0.47 (O~I lines) times 
of that of the \feii\ line at 1.257 $\mu$m.
If the Cas~A ejecta suffered the same extinction $A_{\rm V}$ $\simeq$ 13 of G11.2--0.3,
the \sii\ lines would be $\sim$ 10 times brighter than the \feii\ line at 1.257~$\mu$m,
while the \si\ and O~I lines would be roughly as bright as the \feii\ line,
indicating that the dense \feii\ knots in \gele\ are different 
from the O-rich ejecta found in other SNRs such as Cas~A.
The absence of He~I line (1.083 $\mu$m) in the spectrum of the knot~A 
is also consistent with its ejecta origin, not the circumstellar origin.
The upper limit of the He~I line is almost the same as that of the \si\ line.
The inset in Fig.~\ref{fig_subaru} shows a magnified view of the \feii\ 1.644 $\mu$m line emission,
identifying a weak velocity component at the longer wavelength buried in the main component.
The bottom panel presents a bow shock-like P-V diagram 
of the 1.644 $\mu$m line emission: 
the weak velocity component in the line profile appears as the redshifted emission 
at the both edges compared to the peak emission at the center in the P-V diagram.

The electron number densities of the knot~A and the NW filament
based on the intensity ratios of the \feii\ 1.644 and 1.599 $\mu$m transitions are 
$\sim$ 3 $\times$ 10$^4$ and $\sim$ 5 $\times$ 10$^3$ \cms, respectively.
(Note that it is $\sim$ 7 $\times$10$^3$ \cms\ for the SE filament [Paper I].)
For this we used the atomic parameters from CLOUDY \citep[][]{fet98}
and solved the rate equations of 16 levels.
We also corrected for the extinction effect of $A_{\rm V}$ = 13~mag 
and assumed the electron temperature to be 5,000~K.
The observed (= reddened) fluxes of the 1.644~$\mu$m line emission of the five ejecta knots (A--E) 
are in the range of (3.8--9.0) $\times$ 10$^{-14}$ ergs sec$^{-1}$ cm$^{-2}$, which,
together with the above electron number density of the knot~A,
corresponds to the mass of the ejecta knots in the range of (0.8--1.8)~$\times$ 10$^{-6}$ $M_{\odot}$.

\section{Discussions and Conclusions}

The \feii\ emission in \gele\ primarily consists of 
two bipolar filaments in the NW and SE boundaries and clumpy internal knots 
scattered across the SNR.
In Paper~I we proposed an interpretation that the internal knots and the NW filament are Fe ejecta from the SN explosion, 
while the SE filament has contributions from both the Fe ejecta and swept-up circumstellar material (CSM).
Our results here reaffirm the previous interpretation and also provide new information as follows.
First, the large ($\simgt$~1,000 \kms) blueshifts of the two eastern knots (\ab)
evince that they are indeed Fe ejecta produced in the SN explosion.
Their radial velocities are roughly comparable to those of the SN
ejecta knots in other young SNRs \citep[e.g.,][]{ghw05}, 
while they move much faster than shocked CSM \citep[e.g.,][]{fesen01}.  
The lack of strong emission from any elements other than 
Fe is consistent with the interpretation of the SN ejecta origin of \ab\ as well.
The ejecta nature of \ab\ then asserts that other Fe knots are SN ejecta, including \cde\ in the west.
For the NW and SE filaments,
the former shows the large redshift of $\simgt$ 200~\kms, which, 
together with its knotty emission feature along the inner boundary of the SNR shell,
supports the interpretation that it is an aggregate of Fe SN ejecta knots.
The absence of any significant Doppler shift in the SE filament, on the other hand,
confirms its location at the tangential boundary of \gele,
and the detection of the He~I (1.083 $\mu$m) line emission (see \S~2) manifests
the existence of the swept-up CSM in the SE filament in addition to the Fe ejecta.

The most conspicuous feature in the radial velocity distribution of the Fe ejecta in \gele\ is
the large ($\simgt$~1,000~\kms) differences between the eastern (\ab) and western (\cde) knots
which are projected to be close to the center.
Obviously the large blueshifts of A,~B are quite notable; however,
the small Doppler shifts of \cde\ are also rather unexpected,
given that even the NW filament near the SNR boundary shows more substantial radial velocities.
We briefly consider two possibilities as the origin of the velocity distribution of the knots.
First, the small Doppler shifts of \cde\ can be explained if they have been 
significantly decelerated compared to \ab\ since the SN explosion.
The circumstellar material from the progenitor or material from the outer envelop
of the SN can in principle be responsible for the deceleration.
Secondly, if there is a systematic velocity shift of $\sim$~--500~\kms\ 
between the eastern and western knots, 
their velocity discrepancy can be explained as \ab\ move towards us while \cde\ move away from us.
Such a systematic velocity of the SN ejecta is not rare and
has been reported in several other young SNRs \citep[e.g.,][]{ghw05}.
Or, for the case of \gele, if the central ejecta have been swept up by the central PWN,
the latter may produce such a systematic velocity \citep[e.g.,][]{vet04}.
We note that the bow-shock like P-V diagram of the knot A can be explained by this scenario:
if the knot A has moved towards us, the ram pressure-conserving reverse shock 
of the SNR can produce such a bow shock feature.

No other lines are identified from the Fe ejecta in \gele\ 
than the He~I line from the SE filament. 
This supports the interpretation that the SE filament, in contrast to the other 
knots and filament, has contributions from both the SN ejecta and shocked CSM (Paper~I).
While the detection of the He~I line supports for the existence of the CSM,
the increased abundances in the best-fit spectrum of the Chandra X-ray data \citep{ret03}
may reaffirm the existence of the ejecta in the SE filament,
although the X-ray spectrum appears to be inadeqaute for thorough abundance analyses.
On the other hand,
the absence of S (both S~I and S$^+$) and O~I in our Subaru spectrum
suggests that the Fe ejecta in \gele\ can be much different from those found 
in O-rich SNR Cas~A (\S~2) in their chemical compositions and/or ionization states.
Given the ionization potentials of 
Fe~I (7.90 eV), Fe$^+$ (16.18 eV), S$^{+}$ (23.33 eV), and O~I (13.62 eV),
it is highly unlikely that S is in S$^{++}$ while Fe is in Fe$^+$,
although some O could in principle be O$^+$. 
The absence of S indicates that the Fe ejecta in \gele\ may have little 
contribution from other heavy elements of the SN nucleosysnthesis,  
suggesting that at least part of Fe ejecta in core-collapse SN explosion
can be produced without intense microscopic material mixing.
One way to produce such Fe ejecta in the explosive nucleosynthesis
of the core-collapse SNe can be the \arich\ process where the
high temperature and relatively low density of the shocked innermost 
layers of a SN exclusively produce the Fe \citep{wf91,mcc98,hl03}.
Therefore, the Fe ejecta in \gele\ may represent the end product of 
the explosive SN nucleosynthesis via the \arich\ process
at the progenitor's core which were expelled outside without significant material mixing.

\begin{figure}[htf]
\plotone{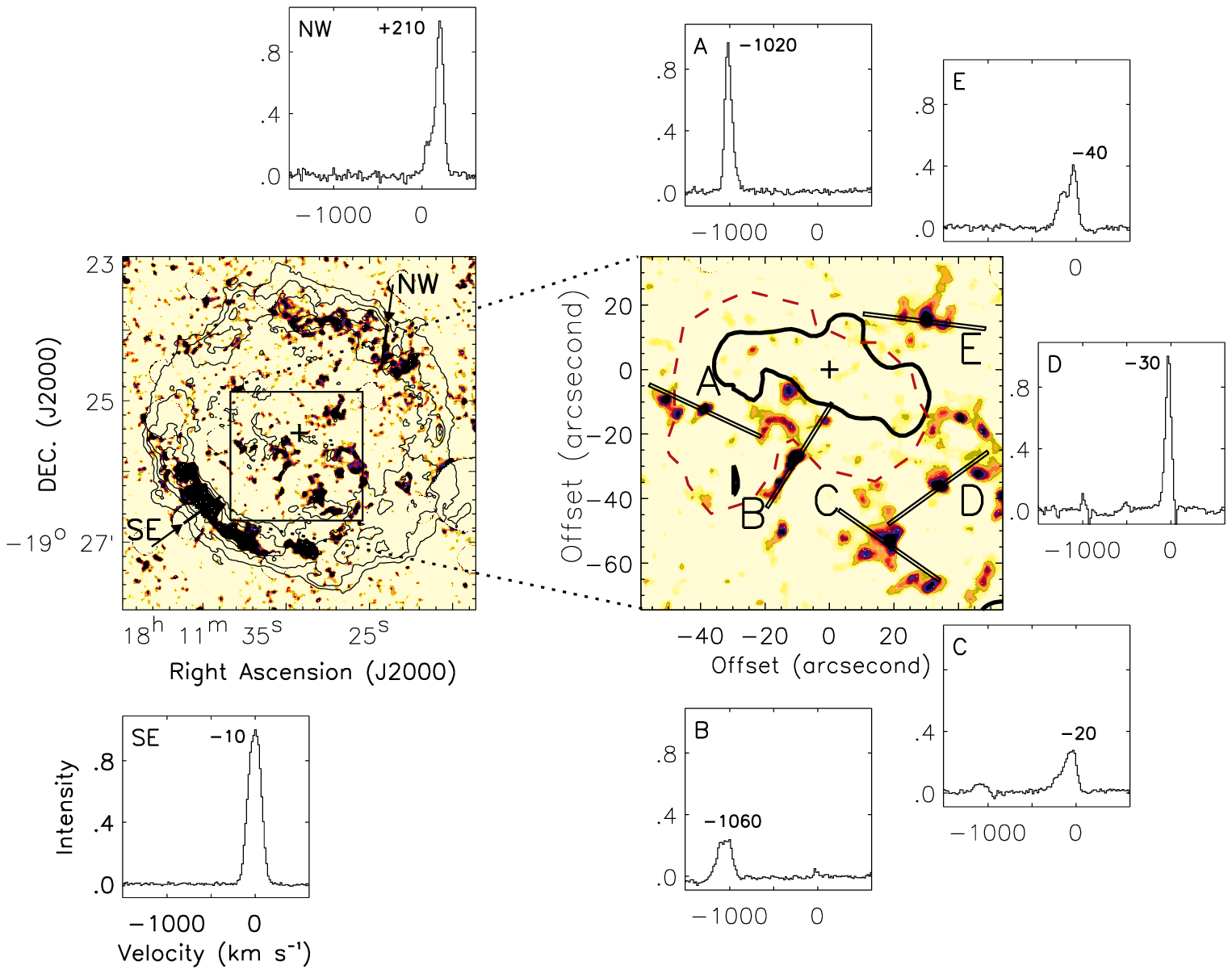}
\caption{
Palomar high-resolution \feii\ 1.644~$\mu$m line spectra of the five knots 
(A--E) and the NW and SE filaments of \gele.
The values near the spectrum peak represent the peak emission velocities.
The velocities are after the subtraction of the +45 \kms\ systematic velocity of \gele\ \citep{get88}.
The intensities (i.e., y-axis) of the five knots are normalized by
the peak intensity of the knot~D, while those of the NW and SE filaments are
normalized by their own peak intensities.
The central images present star-subtracted \feii\ 1.644~$\mu$m emission of \gele\ (Paper~I),
together with locations of the knots and filaments as well as the slit positions.
In the left panel the contours represent radio continuum emission,
and the central cross corresponds to the pulsar J1811--1925 at
(R.A., decl.) = ($\rm 18^h11^m29.22^s$, $-19^\circ25'27.6''$) \citep[J2000;][]{ket01}.
The two arrows point to the slit positions of the NW and SE filaments.
The image in the right panel presents a magnified view of the \feii\ 1.644~$\mu$m emission around the center
along with slit positions of the five knots.
The thick solid contour represents the elongated X-ray emission of the PWN;
the dashed red contour does the boundary the PWN determined by radio tomography.
}
\label{fig_overall}
\end{figure}

The prominent mode of the distribution of the dense Fe ejecta in \gele\ 
is the bipolar pattern along the NW--SE direction of the two filaments.
This direction is almost perpendicular to the NE--SW elongation direction 
of the X-ray PWN (Fig.~\ref{fig_overall}), and the latter direction most likely represents either
the rotational (= polar) or equatorial axis of the pulsar \citep{ket01}.
It is now generally accepted that the core-collapse SN explosion is mostly asymmetric
with major explosion along the progenitor's rotational axis \citep[e.g.,][]{jet05,bet05,let06,wet08},
which predicts the Fe ejecta from the core to mainly distribute in a bipolar pattern 
along the rotational axis. 
Therefore, it may be the case that the NW--SE direction of the Fe ejecta 
represents the progenitor's rotational axis along which the SN primarily exploded.
If so, the NE--SW elongation of the X-ray PWN may correspond to equatorial flows 
from the pulsar as observed around the Crab pulsar \citep{wet00}.
The collimation angles of the NW and SE filaments are somewhat large ($\sim$~60$^\circ$);
however, this is consistent with some predictions of 
core-collapse SN explosions induced by neutrino heating or even magneto-hydrodynamics jets.
For the latter case, it is required that the progenitor is possessed of normal (or weak)
magnetic field and slow rotation to have large collimation angles, 
which agrees with the inferred magnetic field ($\sim$ 1.7~$\times$~10$^{12}$~G) and 
observed rotational period ($\sim$ 65 ms) of the central pulsar of \gele\ \citep{tet99}.

\acknowledgments
We thank Roger Chevalier, Adam Burrows, Ken'ichi Nomoto, Marten van Kerkwijk,
Samar Safi-Harb, Chris Matzner, Andy Howell, and Alex Conley for helpful comments.
D.-S.M. acknowledges the support by NSERC through Discovery program 327277
and B.-C.K. and D.-S.M acknowledge KOSEF through the Joint Research Project
under the KOSEF-NSERC Cooperative Program F01-2007-000-10048-0.

\clearpage

\clearpage

\clearpage

\clearpage
\begin{figure}[htf]
\plotone{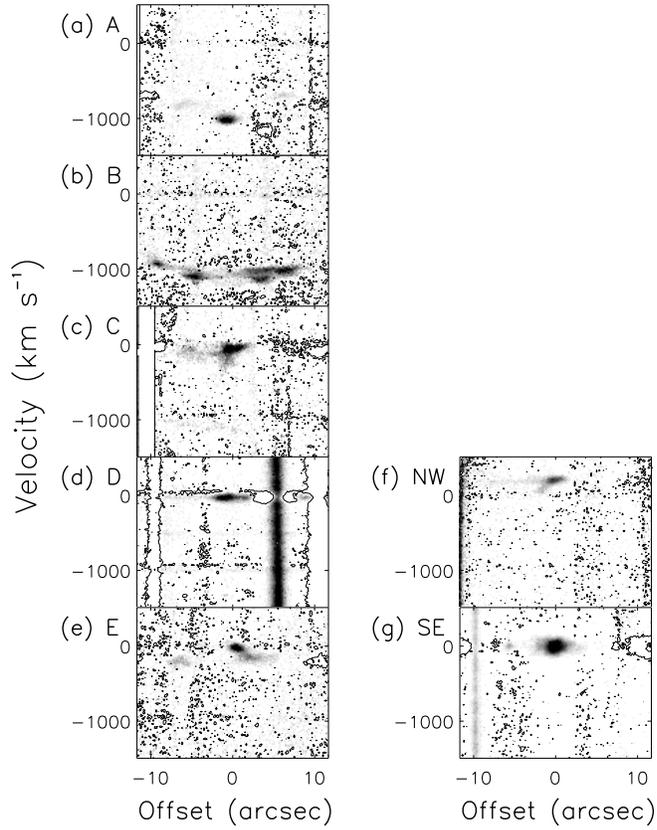}
\caption{
(a)--(g): Same as Fig.~\ref{fig_pv}, but with the velocity coverage of --1400 and 400 \kms.
Also the contours represent the negative pixels to highlight the OH line residuals
resulting from the subtraction of a dithered frame,
while the grey-scale images represent the Fe emission as in Fig.~\ref{fig_pv}.
Note that the OH line residuls around
0 and/or --1000 \kms\ are visible in some panels, especially in (d).
}
\label{fig_pv2}
\end{figure}

\clearpage

\end{document}